\documentclass[superscriptaddress,aps,prb,twocolumn,longbibliography,english]{revtex4-1}
\usepackage{graphicx}
\usepackage{bm}
\usepackage{amsmath,amssymb,amsfonts}
\usepackage{epsfig}
\usepackage{epstopdf}
\usepackage{dcolumn}
\usepackage{grffile}
\usepackage{verbatim}
\usepackage{mathrsfs}
\usepackage{textcomp}
\usepackage[english]{babel}
\usepackage{xr}
\usepackage{nth}
\usepackage[normalem]{ulem}
\usepackage[colorlinks=true,linkcolor=blue,citecolor=blue, urlcolor=blue]{hyperref}
\def\bk{{\mathbf{k}}}
\begin{document}
\title{Electronic Structure of $A$V$_3$Sb$_5$ Kagome Metals}

\author{Keyu Zeng}
\affiliation{Department of Physics, Boston College, Chestnut Hill Massachusetts 02467, USA}
\author{Zhan Wang}
\affiliation{Department of Physics, Boston College, Chestnut Hill Massachusetts 02467, USA}
\author{Kun Jiang}
\affiliation{Beijing National Laboratory for Condensed Matter Physics and Institute of Physics, Chinese Academy of Sciences, Beijing 100190, China}
\author{Ziqiang Wang}
\email{wangzi@bc.edu}
\affiliation{Department of Physics, Boston College, Chestnut Hill Massachusetts 02467, USA}

\date{\today}    

\begin{abstract}
The kagome metals $A$V$_3$Sb$_5$ ($A=$ K, Cs, Rb) have become a fascinating materials platform following the discovery of many novel quantum states due to the interplay between electronic correlation, topology, and geometry. Understanding their physical origin requires constructing effective theories that capture the low-energy electronic structure and electronic interactions. Here, we point out the unusual and puzzling properties of the DFT electronic structure, including the sublattice type of the van Hove singularities, the geometric shape of the Fermi surface, and the orbital content of the low-energy band dispersion, which cannot be described by the commonly used one-orbital or multiorbital kagome tight-binding models. We address these fundamental puzzles and develop an extended Slater-Koster formalism that can successfully resolve these issues. We discover the important role of site-symmetry and interorbital hopping structure and provide a concrete multiorbital tight-binding model description of the electronic structure for $A$V$_3$Sb$_5$ and the family of ``135'' compounds with other transition metals.
\end{abstract}
\maketitle

\section{Introduction}
      The recently discovered vanadium-based kagome metals $A$V$_3$Sb$_5$ ($A=$ K, Cs, Rb) \cite{ortiz_new_2019,ortiz_cs_2020} exhibit a wide range of exotic electronic phenomena. Starting from the high-temperature paramagnetic phase, a host of novel electronic states emerge with a cascade of symmetry breaking in addition to a charge density wave (CDW) transition and continues through the superconducting (SC) transition at low temperatures. These include states breaking crystalline symmetries, CDW and pair density wave (PDW) states \cite{li_observation_2021,chen_roton_2021,nie_charge-density-wave-driven_2022,han_emergent_2024} breaking translation symmetry, electronic nematic phase breaking rotation symmetry \cite{2022Xu}, and the smectic CDW \cite{li_rotation_2022}, charge stripes \cite{zhao_cascade_2021}, and SC state that break both translation and rotation symmetries \cite{chen_roton_2021,deng_chiral_2024}. Despite being non-magnetic and exhibiting the Pauli susceptibility above the CDW transition, there is evidence that the time-reversal symmetry is likely broken in the CDW state \cite{jiang_unconventional_2021,shumiya_intrinsic_2021,xing_optical_2024,mielke_time-reversal_2022,2022Wu,2022Xu,2023Farhang,2023Saykin,2022Guo}, as well as in the SC state \cite{mielke_time-reversal_2022,graham_depth-dependent_2024,le_superconducting_2024,deng_chiral_2024}. In addition to the PDW and the pseudogap behaviors \cite{chen_roton_2021}, unprecedented evidence for charge-$6e$ and charge-$4e$ flux quantization and higher-charge superconductivity has been observed in thin-flake CsV$_3$Sb$_5$ ring devices \cite{ge_charge-4e_2024}.%,zhou_chern_2022,varma_2023}. 
      
      The simplest theoretical model for understanding these correlated phenomena is a one-orbital tight-binding (TB) model on the kagome lattice that includes various electron-electron interactions at band-filling close to the van-Hove singularity (vHS) \cite{wang_competing_2013,kiesel_sublattice_2012,kiesel_unconventional_2013,yu_chiral_2012}. 
      It was found that the $2\times2$ bond-order dominated charge density wave (CDW) can emerge due to the sublattice polarization of the pure-type (p-type) vHS, including the possible time-reversal symmetry breaking loop-current order and unconventional superconductivity \cite{zhou_chern_2022,ferrari_charge_2022,tazai_charge-loop_2023,park_electronic_2021,lin_complex_2021,wu_nature_2021,feng_chiral_2021,denner_analysis_2021,2023Dong,varma_2023,fu_exotic_2024}. 
      
      The kagome metals are multiorbital materials that contain multiple Fermi surfaces (FS). In order to start from a more realistic electronic structure and explore the correlation effects, it is important to construct a multiorbital TB model for the low-energy band structure formed by different orbitals. Indeed, several studies have shown that the interplay between the multiple vHS of different wavefunction symmetries can play an important role in the formation of correlated electronic phases \cite{kang_twofold_2022-1,jeong_crucial_2022,li_origin_2023,li_intertwined_2024,christensen_loop_2022,scammell_chiral_2023}. However, the multiorbital TB model description of the kagome metal is highly challenging. As we will show in the work, the most prominent kagome band features predicted by the DFT, which are considered the basis for an one-orbital model description, turn out to escape such modeling because of the sublattice polarization of the double-vHS and the orientation of the FS. Moreover, the multiorbital contents of the wavefunctions in the band states cannot be captured faithfully by fitting the low-energy band dispersions using the conventional Slater-Koster (SK) formalism \cite{slater_simplified_1954}. The focus of this work is to resolve these difficulties and obtain a faithful multiorbital tight-binding model description of the DFT electronic structure for kagome metal $A$V$_3$Sb$_5$.

      \begin{figure*}[ht]
	\centering  
	\includegraphics[width=\textwidth]{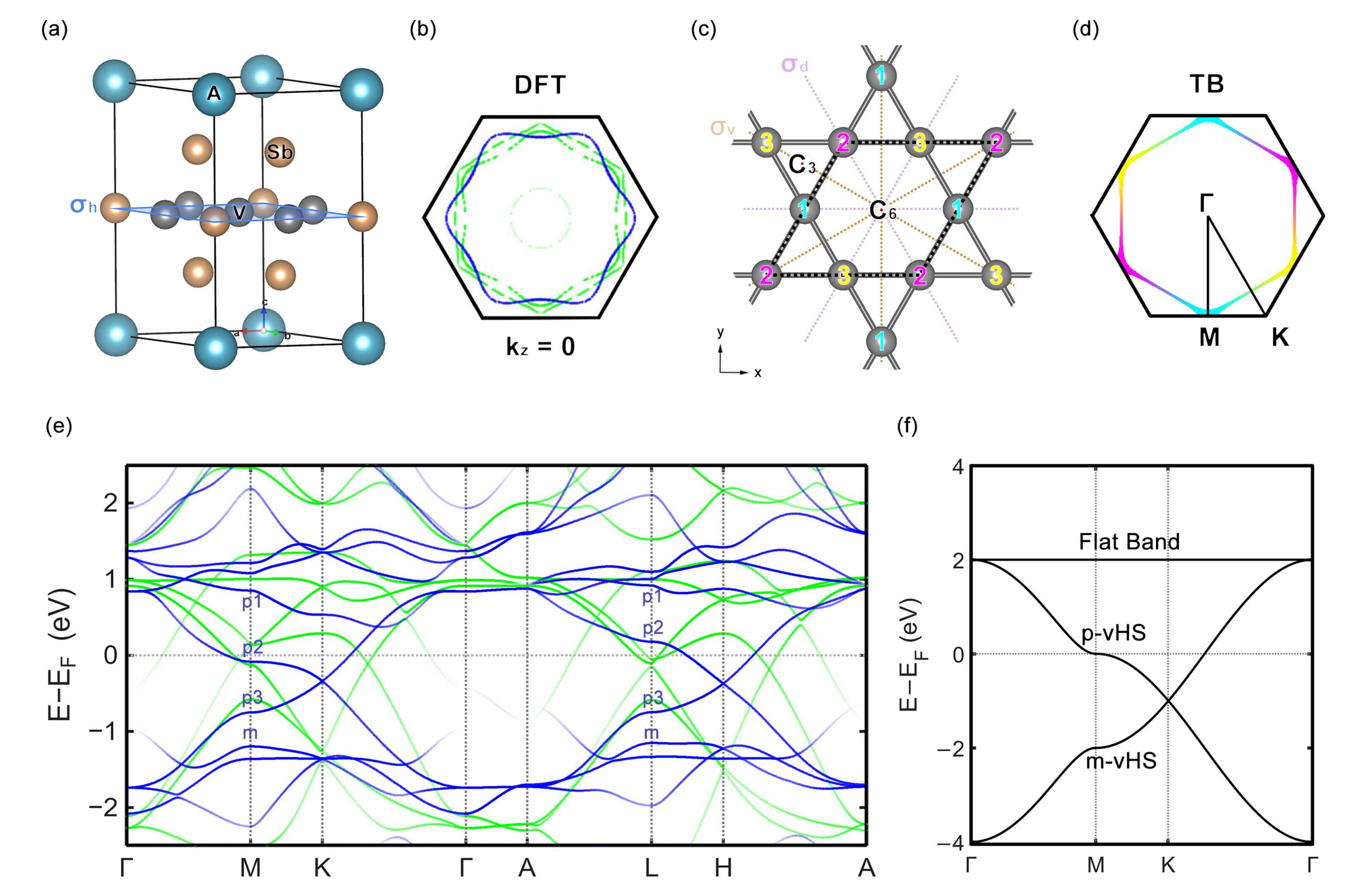}
	\caption{Crystal structure and DFT band structure of $A$V$_3$Sb$_5$. (a) Crystal Structure of $A$V$_3$Sb$_5$ with mirror plane $\sigma_h$ along the kagome plane. (b) The Fermi surfaces and (e) 3D DFT band structure for $\sigma_h$ mirror odd $d_{xz}, d_{yz}$ orbitals (green line) and $\sigma_h$ mirror even $d_{xy}, d_{x^2-y^2}$ and $ d_{z^2}$ orbitals (blue line) on the V atoms of CsV$_3$Sb$_5$. The 3D Brillouin Zone and high-symmetry path is shown in Appendix. (c) The single-layer 2D kagome lattice made of V atoms with important symmetry elements indicated. (d) The Fermi surface and (f) band structure of the one-orbital model with $t=-1.0$ eV, $\mu=0.0$ eV, where the contents of different sublattices are marked in cyan, magenta and yellow.}
	\label{fig1}
     \end{figure*}
      
      Our objective is to elucidate the sublattice and symmetry characteristics of the multiorbital wavefunctions in addition to the low-energy dispersions associated with the $d$-bands of kagome metals. We begin by discussing the fundamental puzzles in the electronic structure. 
      We then develop an extended SK method for constructing multiorbital tight-binding model that has maximum freedom to take into account the crystal fields environment created by Sb atoms while maintaining the $d$-orbital wavefunction symmetry. Using the extended SK model, we successfully capture the shapes of the FSs, the crucial sublattice and orbital contents of the vHS, and the multiorbital contents of the low-energy band dispersion. We find that the interorbital hopping structure between mirror-even and odd $d$ orbitals plays a fundamental role, leading to interorbital flat bands as a key ingredient to understand double p-type vHS (p-vHS) with different orbital symmetries \cite{zeng_interorbital_2024}. We provide a complete tight-binding model and the hopping parameters for the five $d$ orbitals that faithfully accounts for the band structure of the kagome metals. Our theoretical formalism not only describes the electronic structures of $A$V$_3$Sb$_5$ \cite{ortiz_cs_2020,tan_charge_2021,kang_twofold_2022-1,nakayama_multiple_2021,luo_electronic_2022,hu_rich_2022}, CsTi$_3$Bi$_5$ \cite{yang_observation_2023,hu_non-trivial_2023,yang_superconductivity_2024} and CsCr$_3$Sb$_5$ \cite{liu_superconductivity_2024}, but also applies as a universal approach to build compact yet realistic tight-binding models for other transition metal systems.
      
\section{Fundamental Puzzles of the Electronic Structure}
    
    The band structure generated by the one-orbital TB Hamiltonian $H=t\cdot H^K+\mu I$ is shown in Fig~\ref{fig1}(f), where
    \begin{equation}
    H^K(\mathbf{k}) = 2\left[\begin{array}{ccc}
    0 & \mathbf{c}(\mathbf{k}\cdot \mathbf{r}_{21}) &  \mathbf{c}(\mathbf{k}\cdot \mathbf{r}_{31})\\
    \mathbf{c}(\mathbf{k}\cdot \mathbf{r}_{12}) & 0 &  \mathbf{c}(\mathbf{k}\cdot \mathbf{r}_{32})\\
    \mathbf{c}(\mathbf{k}\cdot \mathbf{r}_{13}) & \mathbf{c}(\mathbf{k}\cdot \mathbf{r}_{23}) & 0 
    \end{array}\right],
    \label{Hk}
    \end{equation}
    $\mathbf{c}(x) \equiv \cos(x)$, and $\mu$ is the chemical potential. In Eq.~(\ref{Hk}), $\mathbf{r}_{ij}$ denotes the hopping vector connecting the sublattice sites $i$ and $j$ in the up triangles in Fig.~\ref{fig1}(c), and the wavevector $\mathbf{k}$ is defined in the first Brillouin zone in Fig.~\ref{fig1}(d). Its characteristic features are the Dirac crossing at the K-point, the flat band (FB) and the two vHS,  a p-type and a mixed type (m-type) at the M-points. The pure/mixed classification describes the sublattice contents of the wavefunction of the electronic states at the van Hove singularities, where "pure" indicates the electronic states are located only on a single sublattice \cite{kiesel_sublattice_2012,kiesel_unconventional_2013,kang_twofold_2022-1}. At the $M_2$ point located at $\bold{k}=(\pi,\pi/\sqrt{3})$, the Hamiltonian \begin{equation}
       H(M_2) = 2t\left[\begin{array}{ccc}
        \mu & 0 & 1\\
        0 & \mu & 0\\
        1 & 0 & \mu 
        \end{array}\right].
        \label{M-matrix}
    \end{equation}
    There is no hopping element between the sublattice site $i=2$ and the other two sites $j=1,3$, leading to a sublattice-polarized eigenstate $\vert\Psi_{\text{p-vHS}}\rangle = [0,1,0]$, i.e., the p-vHS at energy $E_{\text{p-vHS}}=\mu$. The other two eigenstates with mixed sublattices are $\vert\Psi_{\text{m-vHS}}\rangle = [1,0,1]$ for the m-vHS at $E_{\text{m-vHS}} = \mu+2t$ and $\vert\Psi_{\text{FB}}\rangle = [1,0,-1]$ for the flat band state at $E_{\text{FB}} = \mu-2t$ as shown in Fig.~\ref{fig1}(f). 
    The Fermi surface connects the p-vHS at two different M points, as shown in Fig.~\ref{fig1}(d). As a result, the unstable electronic states are localized on two different sublattices, thus favoring a bond-order dominant CDW order \cite{kiesel_sublattice_2012}. 
     
\begin{figure*}[ht]
        \centering
        \includegraphics[width=\textwidth]{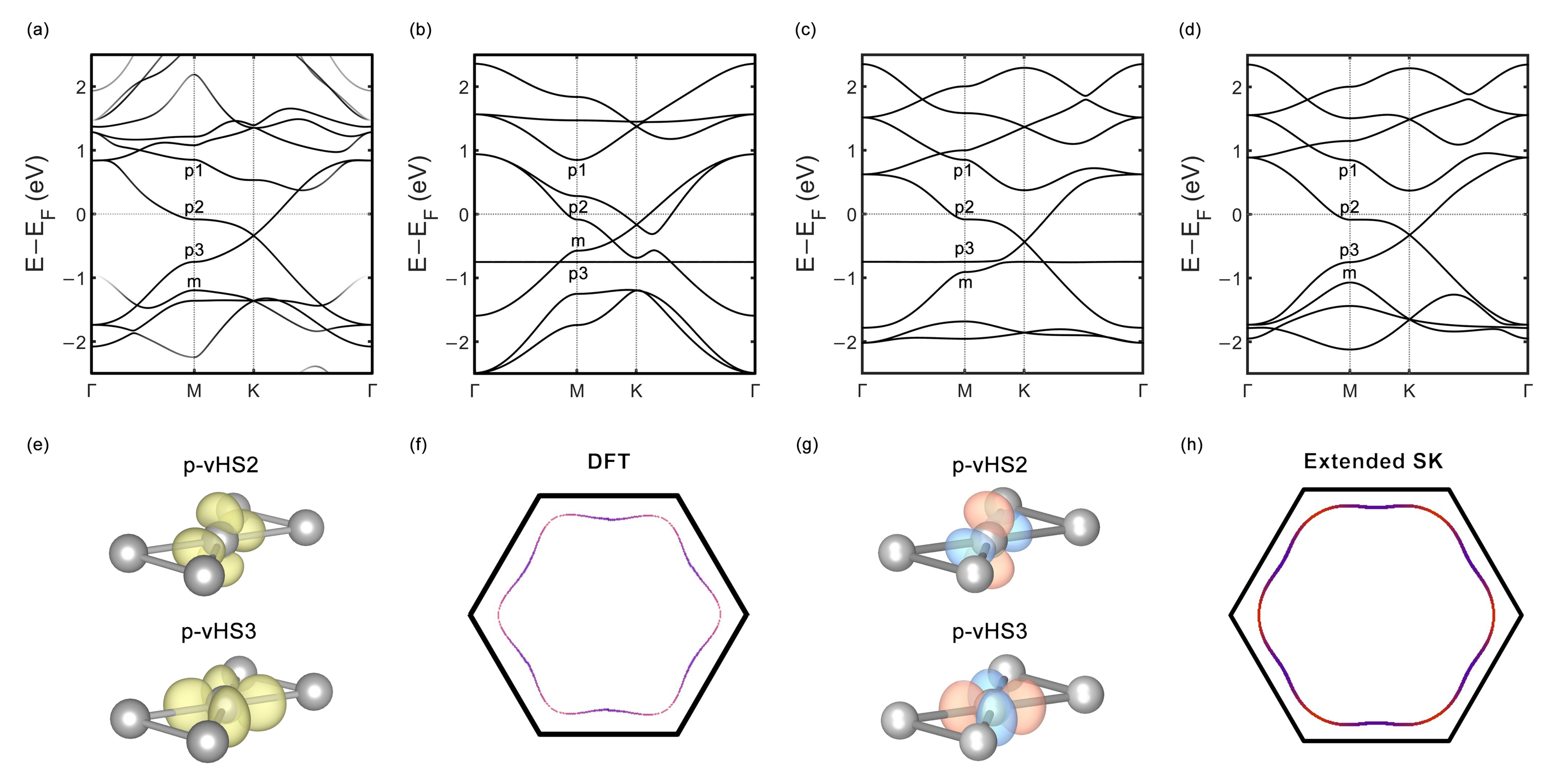}
        \caption{Comparison of DFT calculation and extended SK model results. (a) DFT band structure at $k_z=0$ for $d_{x^2-y^2}, d_{z^2}$ and $d_{xy}$ orbitals on kagome lattice formed by V atoms. Extended SK model results with parameters: $\mu_1 = 0.85, \mu_2 = -0.084, \mu_3 = -0.75$ pinning all three p-vHS. (b) Only \nth{1} nearest neighbor interorbital hopping $t^{23}_\text{\nth{1}} = -0.38, t^{31}_\text{\nth{1}} = -0.35, t^{12}_\text{\nth{1}} = -0.48$ and no intraorbital hopping $t^{\alpha\alpha}_{\nth{1}/\nth{2}}$. (c) Intraorbital hopping for $d_{h_{1/2}}$: $t^{11}_\text{\nth{1}} = -0.05, t^{22}_\text{\nth{1}} = -0.3, t^{11}_\text{\nth{2}} = 0.1, t^{22}_\text{\nth{2}} = 0.2$ and \nth{2} nearest neighbor interorbital hopping $t^{23}_\text{\nth{2}} = 0.12, t^{12}_\text{\nth{2}} = 0.0, t^{31}_\text{\nth{2}} = 0.05$ are added to TB in (b). (d) Intraorbital hopping for $d_{h_{3}}$ $t^{33}_\text{\nth{1}} = -0.05, t^{33}_\text{\nth{2}} = -0.25$ are added to previous parameters. (e) The electron density distribution of the p-vHS2 and the p-vHS3 calculated by DFT. p-vHS1 is not presented due to its relatively high energy. (g) The wavefunction distribution of the p-vHS2 and the p-vHS3 calculated by extended SK model (positive sign in light red and negative sign in light blue). The Fermi surfaces calculated by (f) DFT and (h) the extended SK model with orbital contents marked in red, green and blue for $d^r_{x^2-y^2}$, $d^r_{z^2}$ and $d^r_{xy}$, respectively. Units of parameters are in eV.}
        \label{fig2}
    \end{figure*}

    A natural question arises: can we find the typical one-orbital TB band features in the band structure of realistic materials? The calculated DFT band structure for CsV$_3$Sb$_5$ is shown in Fig.~\ref{fig1}(e) \cite{tan_charge_2021}. Those of the other two members in the $A$V$_3$Sb$_5$ family, as well as for CsTi$_3$Bi$_5$ \cite{yang_observation_2023,hu_non-trivial_2023,yang_superconductivity_2024} and CsCr$_3$Sb$_5$ \cite{liu_superconductivity_2024} are similar. They are in general in good agreement with the band structure measured by ARPES experiments at low energies \cite{kang_twofold_2022-1,nakayama_multiple_2021,luo_electronic_2022,hu_rich_2022}. Surprisingly, there are seemingly one-orbital dispersions formed by the $d_{xy/x^2-y^2/z^2}$ orbitals (marked in blue) in the DFT band structure shown in Fig.~\ref{fig1}(e), namely the two bands producing two vHS (marked as p2 and p3) and the Dirac band crossing at K. However, upon closer inspection, there are crucial differences: (1) the corresponding hexagonal FS plotted in Fig.~\ref{fig1}(b) has a different orientation than the one-orbital prediction (Fig.~\ref{fig1}(d)); (2) More importantly, both vHS are p-type (p2 and p3) in DFT and ARPES instead of a p-vHS and an m-vHS in one-orbital dispersions as shown in Fig.~\ref{fig1}(e). Moreover, both polarization-dependent ARPES measurements and DFT calculation indicate that double p-vHSs are formed by orbitals of different symmetries $\mathbb A_g$ and $\mathbb B_{1g}$, respectively \cite{hu_rich_2022,kang_twofold_2022-1}. Furthermore, the energy of p2 disperses upward along $\bold{k}_z$ to become p2 at the L point in Fig.~\ref{fig1}(e), but the dispersion of p3 is flat along kz and remains approximately unchanged at L. From the electron densities calculated by DFT and shown in Fig.~\ref{fig2}(e), p2 has a substantial distribution along the $z$ direction, hinting at a large content of $d_{z^2}$ orbitals. In contrast, p3 consists mostly of in-plane $d$ orbitals. These multiorbital features of the low-energy electronic structure fundamentally deviate from the one-orbital description, bringing difficulties in accurate electronic modeling of the physical phenomena.

    We propose an extended multiorbital SK model. The comparison between DFT and our results is presented in Fig.~\ref{fig2}(d),(h) and (e).
    
\section{Symmetrized Kagome Orbital Basis}
    \subsection{Orbital symmetry}
        The low-energy electronic structures of kagome metals $A$V$_3$Sb$_5$ are dominated by the $3d$ orbitals of V atoms in the kagome lattice planes with mirror symmetry $\sigma_h$, as shown in Fig.~\ref{fig1}(a). 
        Simplified models with fewer orbitals or an approximation of vanishing interorbital hopping were proposed \cite{hu_rich_2022,okamoto_topological_2022,wu_nature_2021}. There are also models that include additional hopping between V $d$ orbitals and Sb $p$ orbitals \cite{2023Deng,li_origin_2023,li_intertwined_2024} or interlayer hopping with fine-tuned parameters \cite{barman_stacking_2024}. Currently, these models can only partially describe the band dispersion, Fermi surface, and multiorbital wavefunction. 
        Here we focus on the effective electronic model for the five vanadium $3d$ orbitals in a 2D kagome lattice. The orbital contents of the band structure shows a clear separation of the mirror odd orbitals $d_{xz/yz}$ ($\mathbb{E}_{1g}$) and mirror even orbitals $d_{xy/x^2-y^2/z^2}$ ($\mathbb{E}_{2g}$ and $\mathbb{A}_{1g}$) with respect to $\sigma_h$. This is due to the pseudo-2D nature of the kagome lattice planes, which are effectively partitioned by the presence of sizable $A$ cations, as shown in the crystal structure of Fig.~\ref{fig1}(a). The multiorbital TB Hamiltonian of five $3d$ orbitals utilizing the Slater-Koster (SK) formalism \cite{slater_simplified_1954} can also be split into two diagonal blocks: mirror odd $d_{xz/yz}$ and mirror even orbitals $d_{xy/x^2-y^2/z^2}$. 
        
\begin{figure*}[ht]
    	\centering
    	\includegraphics[width=0.8\textwidth]{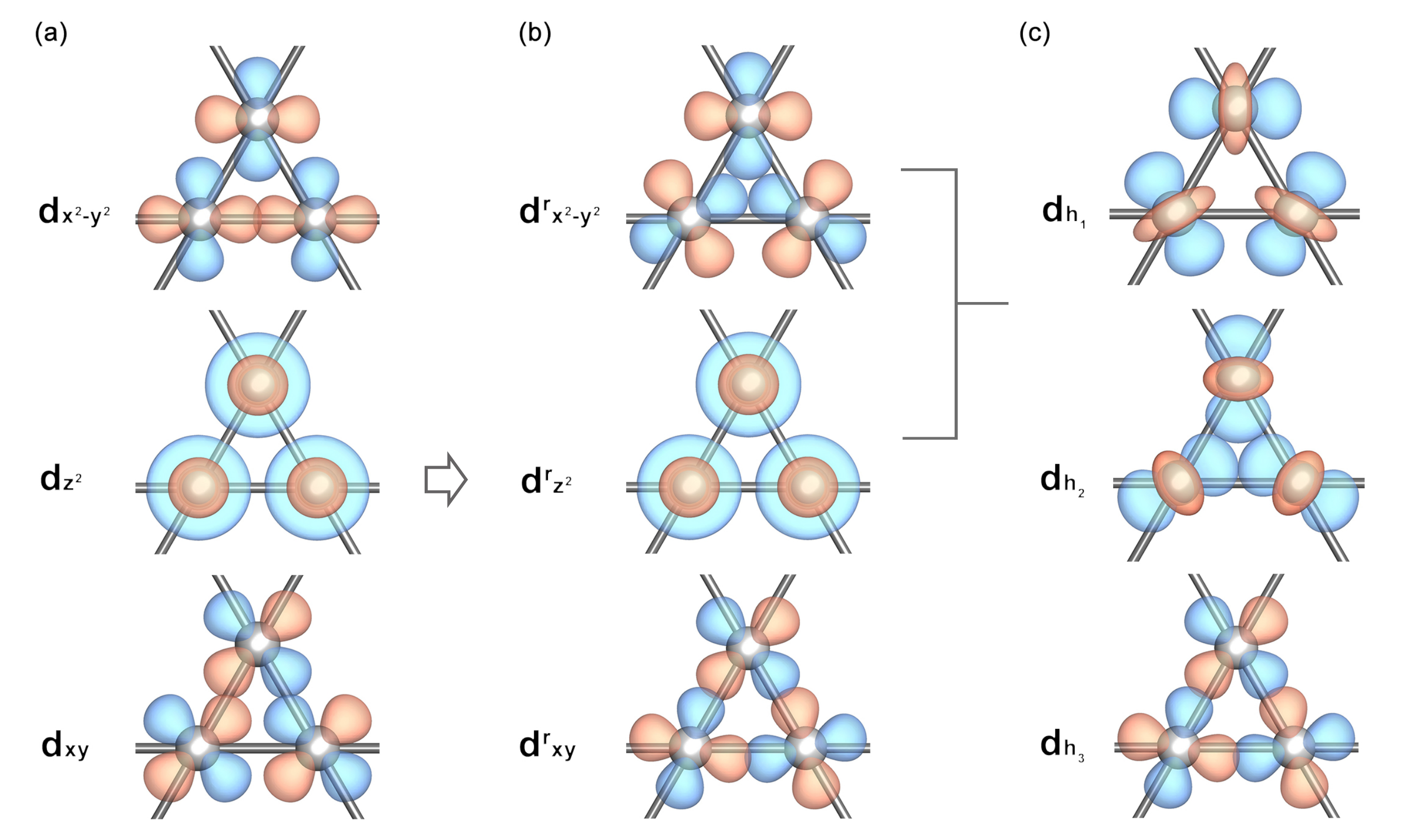}
    	\caption{Change of orbital basis by linear combination. Wavefunction distribution with positive sign marked in light red and negative sign in light blue.} (a) Aligned $d_{x^2-y^2}, d_{z^2}, d_{xy}$ as the standard orbital basis of Slater-Koster TB Hamiltonian. (b) Rotated $d$ orbitals $d^r_{x^2-y^2}$, $d^r_{z^2}$ and $d^r_{xy}$. (c) Hybrid $d$ orbitals with mixing angle $\theta=0.6099=0.2041\pi$ between $d^r_{x^2-y^2}$ and $d^r_{z^2}$.
    	\label{fig3}
        \end{figure*}

        In each symmetry sector, an SK Hamiltonian can be constructed based on the two-center approximation,
        \begin{equation}
            H_{SK} = \sum_{\alpha,\beta}\sum_{\langle ij\rangle}t^{\alpha\beta}_{ij}d^\dagger_{\alpha,i}d_{\beta,j}+\sum_{\alpha}\sum_{i}\mu_{\alpha,i}d^\dagger_{\alpha,i}d_{\alpha,i}
            \label{ham_sk}
        \end{equation}
         where the hopping strength between different orbitals and sublattices $t^{\alpha\beta}_{ij}$ is proportional to the orbital overlapping integral \cite{slater_simplified_1954} with the orbitals aligned on each site as shown in Fig.~\ref{fig3}(a).
        In Eq.~(\ref{ham_sk}), $\langle ij\rangle$ refers to the first and second nearest neighbor hopping between sites on sublattices $i$ and $j$, $\alpha$ and $\beta$ are orbital indices, and $\mu_{\alpha,i}$ is the orbital and sublattice-dependent crystal field. The constant chemical potential is absorbed in $\mu_{\alpha,i}$ for brevity.
        Further decoupling of orbitals based on other in-plane symmetry is not allowed at a general $\mathbf{k}$ point due to its reduced symmetry \cite{slater_simplified_1954}. Thus, each diagonal block is still complicated to understand and compare to the realistic band structure, especially for the mirror even orbitals with three members. 
    
    \subsection{Symmetrization by Rotation}
        Consider first the sector of even orbitals. The SK Hamiltonian in the original orbital basis ($ d_{x^2-y^2},d_{z^2},d_{xy}$) is a complicated $9\times9$ matrix in $\bk$-space,
        \begin{equation}
        H_{e}(\mathbf{k}) = \left[\begin{array}{ccc}
        H_{d_{x^2-y^2},d_{x^2-y^2}} & H_{d_{x^2-y^2},d_{z^2}} & H_{d_{xy},d_{x^2-y^2}} \\
        H_{d_{x^2-y^2},d_{z^2}}^{\dagger} & H_{d_{z^2},d_{z^2}} & H_{d_{xy},d_{z^2}} \\
        H_{d_{xy},d_{x^2-y^2}}^{\dagger} & H_{d_{xy},d_{z^2}}^{\dagger} & H_{d_{xy},d_{xy}}\\
        \end{array}\right],
        \label{T_SKH}
        \end{equation}
        where $H_{\alpha\beta}(\bk)$ describes the hopping between different orbitals on different sublattices with the corresponding crystal field.

        \begin{widetext}
        Since the point group of the kagome lattice has a $C_3$ rotation symmetry (Fig.~\ref{fig1}(c)), it is much more convenient to transform the original orbital basis into one that explicitly respects $C_3$,
        
        \begin{equation}
            \left[\begin{array}{c}
            d^{r\dagger}_{x^2-y^2,i} \\
            d^{r\dagger}_{z^2,i} \\
            d^{r\dagger}_{xy,i} \\
            \end{array}\right] =  R \left[\begin{array}{c}
            d^{\dagger}_{x^2-y^2,i} \\
            d^{\dagger}_{z^2,i} \\
            d^{\dagger}_{xy,i} \\
            \end{array}\right], \quad \quad
            R=\left[\begin{array}{ccc}
            \cos(\theta_{0i}) & 0 & -\sin(\theta_{0i}) \\
            0 & 1 & 0\\
            \sin(\theta_{0i}) & 0 & \cos(\theta_{0i}) \\
            \end{array}\right], \cos(\theta_{0i}) = \mathbf{\hat{r}}_{0i}\cdot \mathbf{\hat{x}}, \sin(\theta_{0i}) = \mathbf{\hat{r}}_{0i}\cdot \mathbf{\hat{y}}
            \label{rot_obt}
        \end{equation}
        where $\mathbf{\hat{r}}_{0i}$ refers to an unit vector from center of a triangle to sublattice i. In the rotated orbital basis ($d_{x^2-y^2}^r, d_{z^2}^r, d_{xy}^r$), the crystal field becomes uniform, i.e. $\mu_{\alpha,i}=\mu_\alpha$, and the SK Hamiltonian in Eq.~(\ref{ham_sk}) transforms according to
        
        \begin{equation}
            H_{SK}^s = RH_{SK}R^\dagger=\sum_{\alpha,\beta}f^{\alpha\beta}(t^\sigma,t^\pi,t^\delta)\sum_{\langle ij\rangle}\left[\epsilon_{\mathbf{\Gamma}(\alpha),\mathbf{\Gamma}(\beta)}+\delta_{\mathbf{\Gamma}(\alpha),\mathbf{\Gamma}(\beta)}\right]d^{r\dagger}_{\alpha,i}d^{r}_{\beta,j}
            +\sum_{\alpha}\mu_{\alpha}\sum_{i}d^{r\dagger}_{\alpha,i}d^{r}_{\alpha,i}
            \label{ham_r}
        \end{equation}
        where $\{\alpha,\beta\}=\{d_{x^2-y^2}^r, d_{z^2}^r, d_{xy}^r\}=\{1,2,3\}$, $\mathbf{\Gamma}(\alpha)$ refers to the irreducible representation (irrep) of orbital $\alpha$, and $\epsilon$ is the Levi-Civita epsilon. 
        In $\mathbf{k}$-space, the symmetrized Hamiltonian has a much simplified structure:
        \begin{equation}
        H^s_{e}(\bk) = \left[\begin{array}{ccc}
        f^{11}H_{11}^K+\mu_1 I & f^{12}H_{12}^{K\dagger} & f^{13}H_{13}^{AK \dagger} \\
        f^{12}H_{12}^{K} & f^{22}H_{22}^K+\mu_2 I & f^{23}H_{23}^{AK\dagger} \\
        f^{13}H_{13}^{AK} & f^{23}H_{23}^{AK} & f^{33}H_{33}^K+\mu_3 I \\
        \end{array}\right],
        \label{sym_HT}
        \end{equation}
        \end{widetext}
        where all the matrix elements $3\times3$ $H_{\alpha\beta}^{K,AK}$ that describe hopping between symmetrized orbitals have precisely the form of the single orbital kagome TB model $H^K$ in Eq.~(\ref{Hk}), or its antisymmetric counterpart,
        \begin{equation}
            H^{AK}(\mathbf{k}) = 2\left[\begin{array}{ccc}
            0 & -\bold{c}(k_3) &  \bold{c}(k_2)\\
            \bold{c}(k_3) & 0 &  -\bold{c}(k_1)\\
            -\bold{c}(k_2) & \bold{c}(k_1) & 0 
            \end{array}\right].
            \label{Hak}
        \end{equation} 
        In Eq.~(\ref{sym_HT}),$f^{\alpha\beta}$ is the hopping amplitude between symmetrized orbitals $\alpha$ and $\beta$, which is now uniform for different combinations of sublattices. They are functions of the SK hopping parameters $f^{\alpha\beta}=f^{\alpha\beta}(t^\sigma,t^\pi,t^\delta)$, and are listed in the Appendix.
        
        This symmetrization process significantly reduces the variety of different hopping strengths, consolidating them into six uniform hopping functions between the sublattices for each combination of orbitals. In effect, this is a process to "kagomize" the complicated multiorbital model in terms of $3\times3$ blocks of the generic one-orbital kagome TB model. Multiorbital Hamiltonians with all near neighbor hoppings can be simultaneously kagomized this way.
        The emergence of the antisymmetric $H_{\alpha\beta}^{AK}$ occurs when the two symmetrized orbitals $\alpha$ and $\beta$ transform differently (i.e. even and odd) under the in-plane mirror operation.
  
        The symmetrizing process can be further elucidated and understood through the lens of group theory. Under point group symmetry $D_{6h}$, the $d_{x^2-y^2}$ and $ d_{xy}$ orbitals belong to the $\mathbb E_{2g}$ irrep and the $d_{z^2}$ orbital belongs to the $\mathbb A_{1g}$ irrep. Rotation based on lattice site representation $\Gamma^{a.s.} = \mathbb A_{1g} +\mathbb E_{2g}$ can be understood as tensor product of the representations \cite{dresselhaus2007group}: $\Gamma_{d_{z^2}^r} = \Gamma^{a.s.}\otimes \mathbb A_{1g}=\mathbb A_{1g} +\mathbb E_{2g}, \Gamma_{d_{x^2-y^2},d_{xy}} = \Gamma^{a.s.}\otimes \mathbb E_{2g}=\mathbb A_{1g} +\mathbb A_{2g} +2\mathbb E_{2g}$, where $\Gamma_{d_{x^2-y^2}^r} =\mathbb A_{1g} +\mathbb E_{2g}$ and $\Gamma_{d_{xy}^r} = \mathbb A_{2g} +\mathbb E_{2g}$. Under site symmetry $D_{2h}$, the irreps of the rotated $d$ orbitals at different sites become the same: $\Gamma^{D_{2h}}_{d_{x^2-y^2}^r/d_{z^2}^r}=\mathbb{A}_g, \Gamma^{D_{2h}}_{d_{xy}^r}=\mathbb{B}_{1g}$. Consequently, the hopping strengths between different sites are unified, and the one-orbital TB Hamiltonian $H^{K}$ for the kagome lattice is recovered for the orbitals belonging to the same irrep. Hopping between orbitals with different symmetry leading to interorbital block Eq.~(\ref{Hak}) is discussed in detail in a recent work \cite{zeng_interorbital_2024} and the following sections. 
        
        \subsection{Crystal Field Effects and Hybrid Orbital Basis}
        %Mixing of $d_{x^2-y^2}^r$ and $d_{z^2}^r$}
        The above model is an effective model for the embedded kagome layer made of V atoms. In realistic materials, other atoms significantly influence the $d$-band structure of the kagome lattice. Studies have found the important role of Sb $p$ orbitals in the band structure \cite{jeong_crucial_2022,li_origin_2023}. Due to the relatively low mixing of the Sb $p$ orbitals and the V $d$ orbitals near the Fermi level, this effect can be well captured by a shift in the crystal field $\mu_{\alpha}$ of the $d$ orbitals. 
        
        A much related aspect is that recent works have demonstrated the importance of interactions between multiple p-vHS with different symmetry properties \cite{li_origin_2023,li_intertwined_2024}. Therefore, determining the p-vHS energies with the correct symmetry properties greatly enhances the accuracy of modeling the electronic structures.
        The energies $E_{p\alpha}$ of the p-vHS for the orbital $\alpha$ are anchored by $\mu_{\alpha}$ because the wavefunctions are the atomic orbitals purely localized at one site. The DFT calculation shows electron density distributions that are different from those of the atomic $d$ orbitals for p-vHS1 and p-vHS2 (Fig.~\ref{fig2}(e)), indicating significant mixing between the rotated $d_{x^2-y^2}^r$ and $d_{z^2}^r$ orbitals.  Therefore, the rotated $d$ orbital basis is still not the proper basis to incorporate the crystal field effects. Mixing between $d_{x^2-y^2}^r$ and $d_{z^2}^r$ is required to produce the correct wave functions of p-vHS1 and p-vHS2. 

        As discussed in section IIIB, the rotated orbitals $d_{x^2-y^2}^r$ and $d_{z^2}^r$ exhibit identical symmetry characteristics under both the point group and the site-symmetry group subsequent to rotation in our theory. Therefore, hybridization between them by any ratio is allowed in our TB model as shown in Fig.~\ref{fig3}(c). Upon the linear combination of $d_{x^2-y^2}^r$ and $d_{z^2}^r$, a new orthonormal hybrid orbital basis $d_{h_{1/2/3}}$ is formed, 
        \begin{align}
              \vert d_{h_1}\rangle &= \cos(\theta)\vert d_{x^2-y^2}^r\rangle-\sin(\theta)\vert d_{z^2}^r\rangle \nonumber\\
              \vert d_{h_2}\rangle &= \sin(\theta)\vert d_{x^2-y^2}^r\rangle+\cos(\theta)\vert d_{z^2}^r\rangle 
              \label{hyb}
        \end{align}
        and $d_{h_3}=d_{xy}$. The Hamiltonian retains the structure of $H^s_{e}$ in Eq.~\ref{ham_r} and Eq.~\ref{sym_HT}, but with a new set of hopping functions $f^{\alpha\beta}_h$.
        This additional degree of freedom represented by the mixing angle $\theta$
        between $d_{x^2-y^2}^r$ and $d_{z^2}^r$ allows the proper modeling of crystal field effects. It will be determined by fitting the TB dispersion to the DFT band structure calculations. We will show that the obtained $\theta$ value allows an addition check of the orbital content of the p-vHS with the DFT electron density distribution. We find an excellent agreement. In previous studies on multiorbital TB models \cite{hu_rich_2022,okamoto_topological_2022,wu_nature_2021}, hoppings between $d_{z^2}^r$ with $d_{xy}^r$ and $d_{x^2-y^2}^r$ orbitals were effectively ignored or eliminated by choice of parameters. In contrast, the current work demonstrates the importance of interorbital physics, which leads to correct modeling of the p-vHS wavefunctions and symmetry properties.
        
\section{The Structure of Multiorbital Hamiltonian}
    Upon selecting a symmetrized orbital basis, our subsequent objective is to manifest a single orbital-like dispersion through the two p-vHS with correct orbital symmetries pinned at the DFT calculated energies. We seek to describe and understand the multiorbital wavefunction properties of the band structure rather than merely fitting the energy dispersions. However, the six different hopping functions $f^{\alpha\beta}$ change when the set of three SK parameters $t^{\sigma/\pi/\delta}$ changes, which makes it difficult to control changes in the band structure. 
    
    \subsection{Relaxation of Parameters}
        Therefore, we extend the SK approach by further relaxing the parameter space from three SK parameters $t^{\sigma/\pi/\delta}$ to six independent hopping parameters $t^{\alpha\beta}$.
        This relaxation process can be understood as the effective kagome lattice $d$-orbital hopping renormalized by the influence of the other atoms in the crystal environment. 
        In the extended SK method, the Hamiltonian for the hybrid orbitals is given in the following form:
        \begin{widetext}
        \begin{equation}
            H_{e} =\sum_{\alpha,\beta}t^{\alpha\beta}\sum_{\langle ij\rangle}(\epsilon_{\mathbf{\Gamma}(\alpha),\mathbf{\Gamma}(\beta)}+\delta_{\mathbf{\Gamma}(\alpha),\mathbf{\Gamma}(\beta)})c^{\dagger}_{\alpha,i}c_{\beta,j}+\sum_{\alpha}\mu_{\alpha}\sum_{i}c^{\dagger}_{\alpha,i}c_{\alpha,i}
            \label{ext_H}
        \end{equation}
        where $\{\alpha,\beta\} =\{d_{h_1}, d_{h_2}, d_{h_3}\}$. In the $\bk$ space, the Hamiltonian can be expressed as a matrix: \begin{equation}
        H_{e}(\bk) = \left[\begin{array}{ccc}
        t^{11}H_{11}^K+\mu_1 I& t^{12}H_{12}^{K\dagger} & t^{13}H_{13}^{AK\dagger} \\
        t^{12}H_{12}^{K} & t^{22}H_{22}^K+\mu_2 I & t^{23}H_{23}^{AK\dagger} \\
        t^{13}H_{13}^{AK} & t^{23}H_{23}^{AK} & t^{33}H_{33}^K+\mu_3 I  \\
        \end{array}\right],
        \label{ext_Hk}
        \end{equation}
        \end{widetext}
        which is formally identical to Eq.~(\ref{sym_HT}), except that $f^{\alpha\beta}$ is replaced by $t^{\alpha\beta}$. With $\mu_\alpha$ fixed by the positions of the three p-type vHS in the DFT band structure, we can tune the one-orbital kagome / antisymmetric kagome eigenvalues in each block and the coupling between different blocks in the Hamiltonian by adjusting $t^{\alpha\beta}$. This approach enhances our understanding of the original SK Hamiltonian and the resulting multiorbital band structure. In contrast to the previous multiorbital models, the extended SK model maximizes the degrees of freedom for orbital mixing and crystal field shifts while maintaining the orbital symmetries.
        
    \subsection{Importance of Interorbital Hopping Structure}
        \begin{figure*}[ht]
	       \centering
    	   \includegraphics[width=\textwidth]{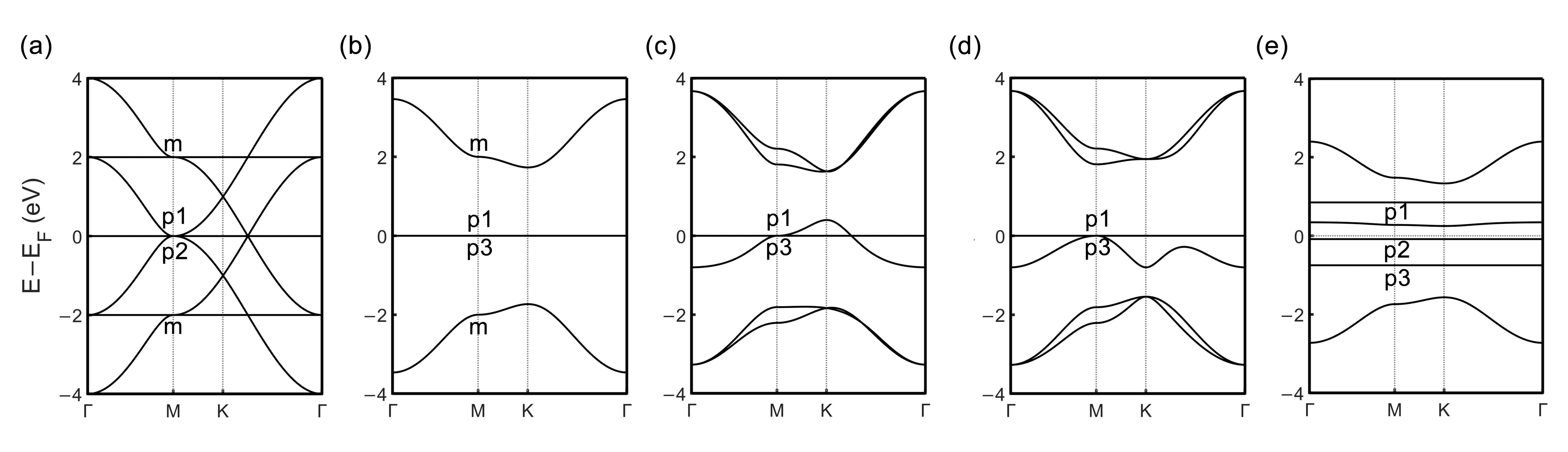}
    	   \caption{Band structure formed by different Hamiltonian blocks. (a) First nearest neighbor (\nth{1} nn) non-zero interorbital $H^K$ hopping with $t^{12}=1.0$ eV with other zero hopping. Three FB due to isolated $d_{h_3}$ orbitals with no hopping exist at $\mu_{3}$. (b) \nth{1} nn non-zero interorbital $H^{AK}$ hopping with $t^{31}=1.0$ eV  with other zero hopping. Two mirror interorbital FB consist of $d_{h_1}$ and $d_{h_3}$ with p-vHS pinned at $E=0.0$ eV. Three FB due to isolated $d_{h_2}$ orbitals with no hopping exist at $\mu_{2}$. (c) \nth{1} nn intraorbital hopping $t^{22}_\text{\nth{1}}$ and (d)\nth{2} nn intraorbital hopping $t^{33}_\text{\nth{2}}$ are added to Hamiltonian in (b) with $t^{31}=1.0$ eV. For (a-d), all crystal fields are set to zero: $\mu_{1}=\mu_{2}=\mu_{3}=0.0$ eV. (e) Different crystal fields are added: $\mu_{1}=0.85$ eV, $\mu_{2}=-0.084 eV, \mu_{3}=-0.75$ eV, with interorbital \nth{1} nn $t^{23}=t^{31}=-0.5$ eV with other zero hopping. The dispersive bands are doubly degenerate.}
    	   \label{fig4}
        \end{figure*}

        The $H^K$ blocks, i.e. $3\times3$ matrix elements containing $H_{\alpha\beta}^K$ in Eq.~(\ref{ext_Hk}), is a universal hopping Hamiltonian for orbitals that share the same symmetry properties under both the point group and the site-symmetry group \cite{zeng_interorbital_2024}. Therefore, each intraorbital block has the same one-orbital $H^K$ structure. As such, the properties of the van Hove singularities, the shape and orientation of the Fermi surface, etc. are identical to the one-orbital kagome model. Among the three interorbital blocks, a $H^K$ block exists between $d_{h_1}$ and $d_{h_2}$ orbitals because they both belong to the irrep $\mathbb{A}_1$. If the other interorbital blocks are zero, there will be one set of antibonding and one set of bonding one-orbital kagome band dispersions with effective hopping $\pm t^{12}$, as shown in Fig.~\ref{fig4}(a). Different $\mu_1 \neq \mu_2$ separate the two sets of bands.
        
        In contrast, the $H^{AK}$ blocks between orbitals of different mirror symmetries, $d_{h_1}$ and $d_{h_3}$, and $d_{h_2}$ and $d_{h_3}$, are real antisymmetric (skew-symmetric) matrices. The properties of $H^{AK}$ and its emergent mirror interorbital flat bands have been studied in detail in a recent work \cite{zeng_interorbital_2024}.
        Let us first consider the case where all intraorbital hoppings $t^{\alpha\alpha}$ are set to zero. Two flat bands, which are composed of pure orbital contents, emerge from the interorbital antisymmetric matrix block $H^{AK}$. The energies of these bands are equal to the crystal fields of the hybrid orbitals $\mu_{1/2/3}$. In Fig.~\ref{fig4}(b), $\mu_{2}=\mu_{3}$ leads to doubly degenerate flat bands and dispersive bands. The energy and wavefunction of the mirror interorbital flat band are completely different from the flat band in the original one-orbital $H^K$ bands. Surprisingly, the eigenstates of the FB at the M point are exactly the p-vHS wavefunction as indicated in Fig.~\ref{fig4}(b). 
        
        Next, consider one of the mirror interorbital flat bands of pure orbital content $\alpha$. When intraorbital hopping between the first or second nearest neighbor is introduced for the orbital $\alpha$, it begins to disperse. However, as discussed in section IIIC, the energy of the p-vHS does not change by the hopping between different sublattices. As a result, the dispersion of the $\alpha$ band resembles the one-orbital dispersion through the p-VHS pinned at $\mu_\alpha$, as depicted in Figs.~\ref{fig4}(c) and (d) for $\alpha=3$. These characteristics of the mirror interorbital flat band provide significant insights into the novel double p-vHS present in the final low-energy electronic structure discussed in the following.

    \section{Understanding the Multiorbital Electronic Structure}
    Having understood the significance of the interorbital hopping, in this section, we obtain the full dispersion spectrum and resolve the puzzles in the multiorbital wavefunctions of the DFT electronic structure (Fig.~\ref{fig2}). 
    We continue to focus the discussion on the sector of the three atomic orbitals $d_{xy}, d_{x^2-y^2}$ and $ d_{z^2}$. The results for the sector of $d_{xz}$ and $d_{yz}$ orbitals are provided in the Appendix section.
    \begin{figure*}[ht]
        \centering
        \includegraphics[width=0.8\textwidth]{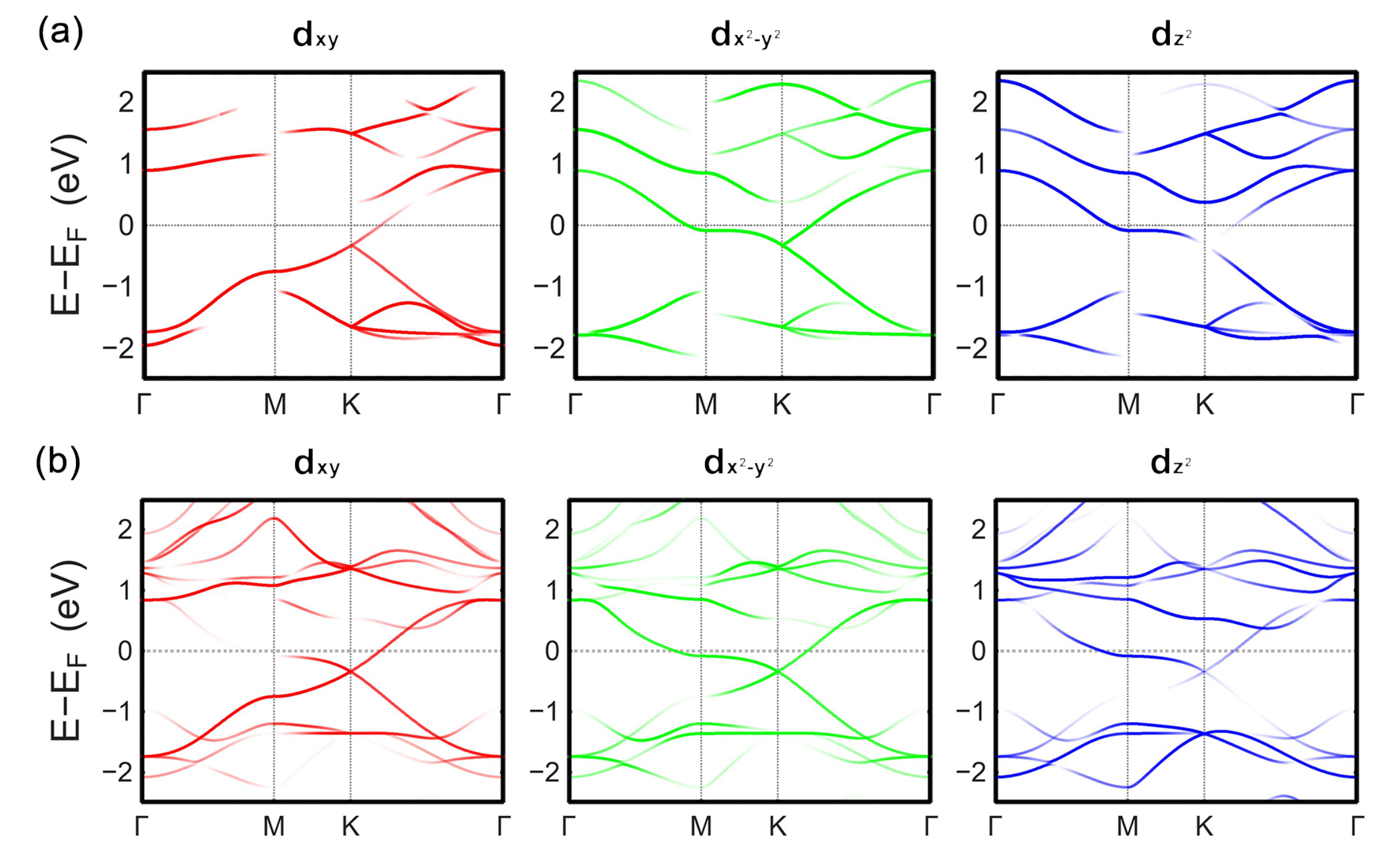}
        \caption{Comparison of orbital contents for DFT calculation and extended SK model. The orbital contents of sublattice-2 (red: $d_{xy}$, green: $d_{x^2-y^2}$ and blue: $d_{z^2}$) of band structure calculated by (a) Extended SK model and (b) DFT. Sublattice-2 is chosen because the p-vHS wavefunctions along the chosen high-symmetry path is localized on sublattice-2.}
        \label{fig5}
    \end{figure*}
    
    \begingroup
        \renewcommand{\arraystretch}{1.5} % Default value: 1
        \begin{table}
            \centering
            \begin{tabular}{|c|c|c|c|c|c|c|c|c|c|} \hline 
                 (eV)& $t^{11}$ & $t^{22}$ & $t^{33}$ & $t^{12}$& $t^{23}$& $t^{31}$& $\mu_{1}$ & $\mu_{2}$ & $\mu_{3}$\\ \hline 
                 $t_\text{\nth{1}}$&  -0.05&  -0.30&  -0.05&  -0.48& 0.38& -0.35& 0.85& -0.084& -0.75\\ \hline 
                 $t_\text{\nth{2}}$&  0.10&  0.20&  -0.25&  0.00& -0.12& 0.05& \multicolumn{3}{|c|}{}\\ \hline
            \end{tabular}
            \caption{Parameters of the band structure calculated by extended SK method for $\{1,2,3\} = \{d_{h_1}, d_{h_2}, d_{h_3}\}$ orbitals.}
            \label{eparams}
        \end{table}
    \endgroup
        
    \subsection{Emergent Flat Band as the Origin of Double p-vHS}
        First, we determine the crystal field $\mu_{1/2/3}$ in the extended SK tight-binding model in Eq.~(\ref{ext_Hk}) from the energies of the p-vHS in the DFT calculations in Fig. 2(a). The $\mu_{1/2/3}$ values are listed in Table ~\ref{eparams}, where $\mu_{3}$ is chosen to be at $E_{p3}=-0.75$ eV based on the symmetry of the $d_{h_3}=d^r_{xy}$ orbital verified by ARPES measurements \cite{hu_rich_2022,kang_twofold_2022-1}. The hopping parameters $t^{\alpha\beta}$ used in our TB model Eq.~(\ref{ext_Hk}) to generate the final band structure in Fig.~\ref{fig2}(d) are also listed in Table ~\ref{eparams}.
        To understand how the final band structure (Fig.~\ref{fig2}(d)) contains a seemingly one-orbital dispersion through double p-vHS, we plot intermediate steps in Figs.~\ref{fig2}(b) and (c).

        When there is no intraorbital hopping and non-zero interorbital hopping $t^{23}$ and $t^{31}$, there are three mirror interorbital flat bands that belong to the three orbitals and are pinned at their crystal fields $\mu_{1/2/3}$, respectively. In addition, there are six dispersive bands, as shown in Fig.~\ref{fig4}(e).
        If comparable $t^{12}$ are added to the Hamiltonian, the flat bands of $d_{h_1}$ and $d_{h_2}$ orbitals hybridize and disperse under symmetric interorbital block $t^{12}H^K$. Only the flat band that is purely composed of orbital $d_{h_3}$ is left and pinned at its crystal field $\mu_{3}$ as depicted in Fig.~\ref{fig2}(b). 
        This serves as a starting point for a complete description of the multiorbital band structure for the kagome metals.
        
        The primary structure of the energy dispersion is then formed by lifting the dispersive band between p1 and p2 in energy by adding intraorbital $t^{11/22}$. 
        Intriguingly, this simultaneously pushes the band passing the m-VHS (m) down in energy, so that the m point now falls below p3 of the mirror interorbital flat band, as shown in Fig.~\ref{fig2}(c).
        Crucially, including the second nearest neighbor hopping in $t^{13/23}$ causes the hybridization and the opening of a hybridization gap between the mirror interorbital flat band and the dispersive band through the m-vHS, in analogy to the hybridization between the flat f-band and the conduction band in heavy fermion compounds.
        As shown in Fig.~\ref{fig2}(c), the  dispersions through p-vHS2 (p2) and p-vHS3 (p3) are now mimicking the one-orbital-like kagome dispersion with a Dirac crossing in between located at the K point. This is the microscopic origin for the formation of double p-vHS in the one-orbital-like kagome dispersion in the ``135'' family of kagome metals, including $A$V$_3$Sb$_5$ and $A$Ti$_3$Bi$_5$.

        The final step is to push down the energy near the center of the Brillouin zone $\Gamma$ point of the flat band passing through p3. This can be achieved by adding the non-zero intraorbital hopping $t^{33}$ between the 1st and \nth{2} nearest neighbors. As discussed in Section IVB, when the non-zero \nth{2} nearest neighbor $t^{33}$ is added, the mirror interorbital flat band of $d_{h_3}$ begins to disperse like in a one-orbital band along $\Gamma$-M with the overall energy lowered. With a relatively large \nth{2} nearest neighbor $t^{33}$ added to the band structure in Fig.~\ref{fig2}(c), the hybridization gap of the flat band increases, pushing down the energy of the band dispersion from $M-\Gamma$ while leaving the p-vHS3 unchanged. This gives the final results of the extended SK multiorbital band structure shown in Fig.~\ref{fig2}(d) obtained with the complete set of parameters provided in Table ~\ref{eparams}.
        We stress that we did not attempt to numerically fit the DFT band dispersion by fine-tuning the parameters. Rather, we focused on resolving the key features in the electronic structure by understanding the structures of the Hamiltonian. After resolving these puzzles, the very good agreement between Fig.~\ref{fig2}(d) and Fig.~\ref{fig2}(a) is remarkable.

    \subsection{Determination of Orbital Contents}
        In our extended SK method, the orbital basis is constructed according to symmetry rather than to the detailed wavefunctions of the orbitals. This enabled free mixing between orbitals of the same symmetry.
        Thus, only the orbital content of $d^r_{xy}=d_{h_3}$, which does not mix with others because of the unique symmetry, is directly determined. The orbital contents of the $d^r_{x^2-y^2}$ and $d^r_{z^2}$ orbitals are, on the other hand, determined by projecting the hopping parameters obtained $t^{\alpha\beta}$ in Table ~\ref{eparams} to the Slater-Koster hopping functions $f^{\alpha\beta}(t^\sigma,t^\pi,t^\delta,\theta)$ (listed in the Appendix), where $\theta$ is the mixing angle of the $d_{x^2-y^2}^r$ and $d_{z^2}^r$ orbitals in Eq.~(\ref{hyb}). In this process, the nonlinear least-squares solver (\emph{lsqnonlin} function of MATLAB) is used to fit the parameters $t_\text{\nth{1}}^{\sigma/\pi/\delta}$ and $\theta$ for the \nth{1} nn hopping. The fitting yields $t_\text{\nth{1}}^\sigma=-0.1855$, $t_\text{\nth{1}}^\pi=0.2735$, $t_\text{\nth{1}}^\delta=0.3091$, and $\theta=0.6099=0.2041\pi$.

        Quite remarkably, the wavefunction distribution obtained for hybrid orbitals for p-vHS2 and p-vHS3 in Fig.~\ref{fig2}(g) agrees with the DFT calculated electron densities in Fig.~\ref{fig2}(e). The mixing angle $\theta_{TB}=0.2041\pi$ is close to the DFT mixing angle $\theta_{DFT}=0.2236\pi$, which serves as a benchmark for the effectiveness of our extended SK method. Furthermore, just as in the electron density distributions calculated by DFT, our wavefunction distributions obtained for the three p-vHS are similar in $A$V$_3$Sb$_5$ and $A$Ti$_3$Bi$_5$, indicating similar crystal field environments for the kagome lattice in this family of kagome metals. 

        Due to the small values of the \nth{2} nn hopping, we do not determine the mixing angle from the \nth{2} nn hoppings separately to avoid large errors. Instead, the above mixing angle $\theta=0.2041\pi$ is used for the linear fit. We obtain the SK parameters for the \nth{2} nn hopping $(t_\text{\nth{2}}^\sigma,t_\text{\nth{2}}^\pi,t_\text{\nth{2}}^\delta)=\left(0.1505, -0.0794, 0.0601\right)$. Note that although the complete set of SK parameters
        $t_\text{\nth{1}}^{\sigma,\pi,\delta}$ and $t_\text{\nth{2}}^{\sigma,\pi,\delta}$ are given, using them in the SK model will not be able to generate a good description of the DFT band structure, signaling the importance of the extended SK method.
        
        The orbital contents of $d_{xy}, d_{x^2-y^2}$ and $d_{z^2}$ are well captured not only for the p-vHS but also at general $\bk$ points, as shown in Fig.~\ref{fig5}. 
        We stress that the success in describing the orbital contents demonstrates the power of understanding the structure of the Hamiltonian and wavefunction properties, rather than just fitting many parameters to the energy dispersion.
        With the well-described multiorbital properties, we can study the electronic instabilities at p-vHS of the real kagome materials. 
    
        \subsection{Fermi Surface and Higher Order vHS}
        The \nth{2} nn hopping between the hybrid orbitals in Eq.~(\ref{ext_H}), i.e. $t_\text{\nth{2}}$, also influences the dispersion around the p-vHS and the shape of FS significantly.
        In the DFT (Fig.~\ref{fig2}(a)) and our multiorbital model (Fig.~\ref{fig2}(b)), the band carrying the p-vHS2 near the Fermi level changes from having a pure $d_{h_2}$ content at the M point to a dispersion with highly mixed orbital contents at the Dirac crossing the K point. Thus, a small change in $t^{23/12}_\text{\nth{2}}$ significantly modifies the band curvature around p-vHS and the energy of Dirac crossing at the K point. This finding is consistent with a recent theoretical work on the important role of \nth{2} nn hopping between V atoms in the formation of CDW states \cite{zhang_atomistic_2024}.
       \begin{figure}
    	\centering
    	\includegraphics[width=\columnwidth]{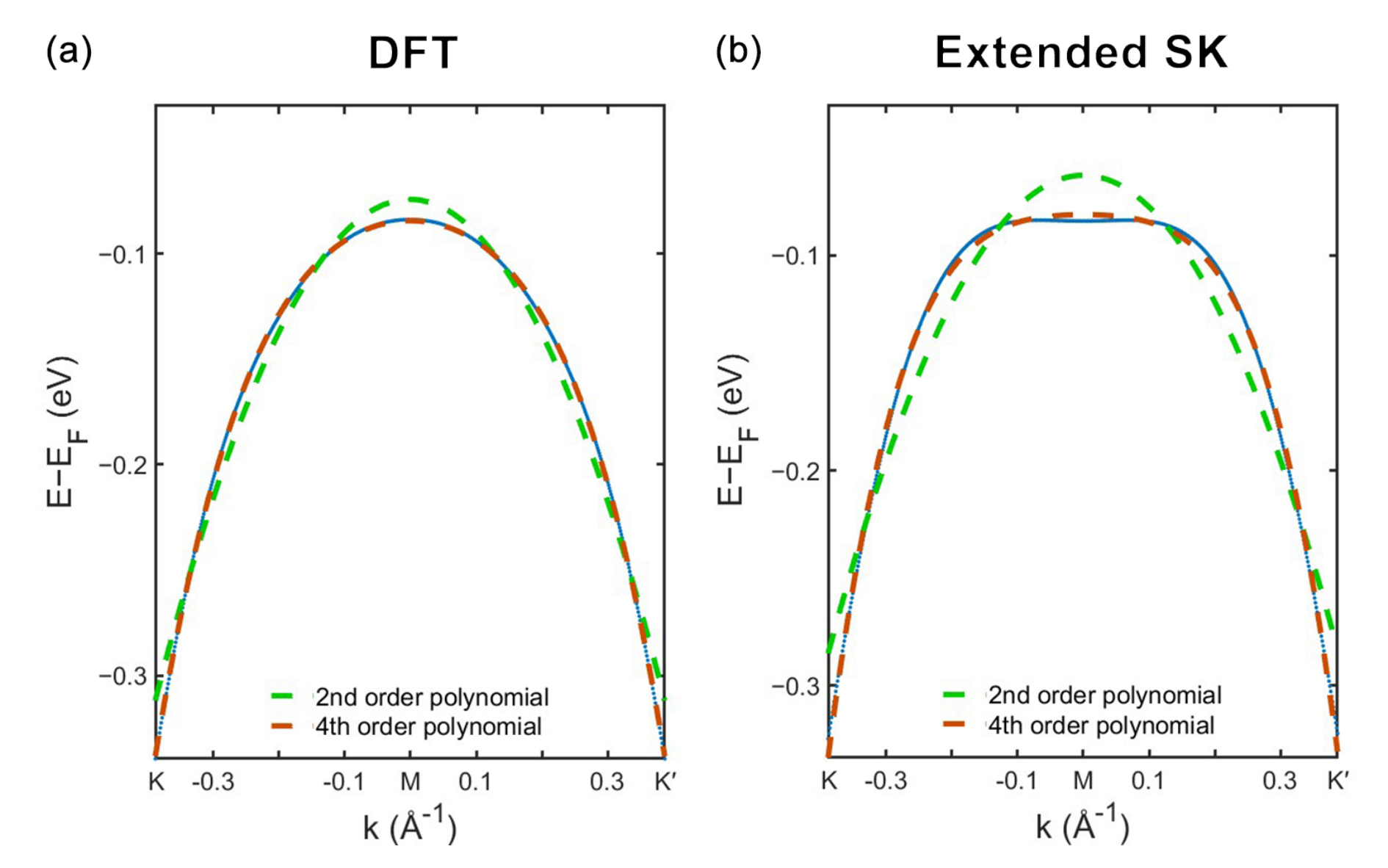}
    	\caption{Higher-order vHS in DFT and extended TB model. The second order and fourth order fitting of the upper vHS dispersion produced by (a) DFT calculation and (b) extended SK model along the K-M-K$^\prime$ path.}
    	\label{fig6}
        \end{figure}
        
        With the proper \nth{2} nn hopping included in the extended SK model, the dispersion around the p-vHS becomes higher order and is better fitted by fourth-order polynomials, as shown in Fig.~\ref{fig6}(b). This feature is consistent with the observations of recent ARPES experiments \cite{hu_rich_2022,kang_twofold_2022-1}.
        Moreover, the rotated FS is well captured in our final results, as shown in Fig.~\ref{fig2}(h) compared to the DFT results in Fig.~\ref{fig2}(g), with matching orbital contents. 
        
        The tight-binding model description of the DFT electronic structure in the $d_{xz}$ and $d_{yz}$ orbital sector can be obtained using the same extended SK formalism, as described in Appendix D. The extended SK model parameters are given in Table ~\ref{oparams}, and the results are presented in Fig.~\ref{fig8}. Both the band dispersions, including the distribution of the vHS, and the FSs show excellent agreement with the DFT results.
       
        In summary, the obtained multiorbital TB model based on the extended Slater-Koster Hamiltonian successfully produces the low-energy band structure for the $d$ orbitals of the V atoms. Not only is the energy dispersion along the high-symmetry path well described but also the orbital contents and sublattice contents of the p-vHSs are accurately captured. This effectively addresses the challenges associated with the one-orbital model and the previous multiorbital models.
        
    \section{Discussion}
        In this work, we develop a systematic approach to constructing an extended Slater-Koster multiorbital TB Hamiltonian. The model allows for maximum orbital mixing and crystal field effects, while maintaining full orbital symmetry. We demonstrate the success of this method in describing the low-energy electronic structure of kagome metals $A$V$_3$Sb$_5$ ($A=$ K, Cs, Rb). The same algorithm can be applied straightforwardly to the other members of the ``135" kagome materials such as CsTi$_3$Bi$_5$ and CsCr$_3$Sb$_5$. Important wavefunction features in addition to the band dispersions are all well captured and thoroughly understood, such as the novel double p-vHS, the rotated Fermi surface, and their sublattice and multiorbital contents. 
        Such a faithful representation of the electronic structure by our extended Slater-Koster TB model can serve as a solid starting point for studying the electronic correlation effects beyond the DFT to achieve more realistic understanding of the correlated charge density wave phases in kagome metals $A$V$_3$Sb$_5$.
        
        Our findings highlight the prevalence of interorbital hopping in realistic materials, which is particularly vital when the properly oriented orbitals exhibit distinct symmetry under the site-symmetry group of the lattice. The application of this realistic, yet concise model to transition-metal systems lays a robust foundation for interpreting experimental band structures. It sets the stage for future exploration of new electronic phases of matter.
        
    \section{Acknowledgement}
        The authors thank Yuxin Wang and Hengxin Tan for valuable discussions. The work is supported by the U.S. Department of Energy, Basic Energy Sciences Grant DE-FG02-99ER45747 and by the Research Corporation for Science Advancement Cottrell SEED Award No. 27856. This work was completed while ZW was on sabbatical leave at the Kavli Institute for Theoretical Sciences (KITS), University of the Chinese Academy of Sciences. ZW thanks KITS for hospitality. 
    
    \section{Reference}
    \bibliography{main}

    \section{Code and Data Availability}
    The code generated for this project is publicly available in the Julia programming language as a versatile package for general lattices and orbitals \cite{MulOrbTB2025}. 

    \appendix
    \section{DFT Calculations}
        \begin{figure}[ht]
    	\centering
            \includegraphics[width=0.5\columnwidth]{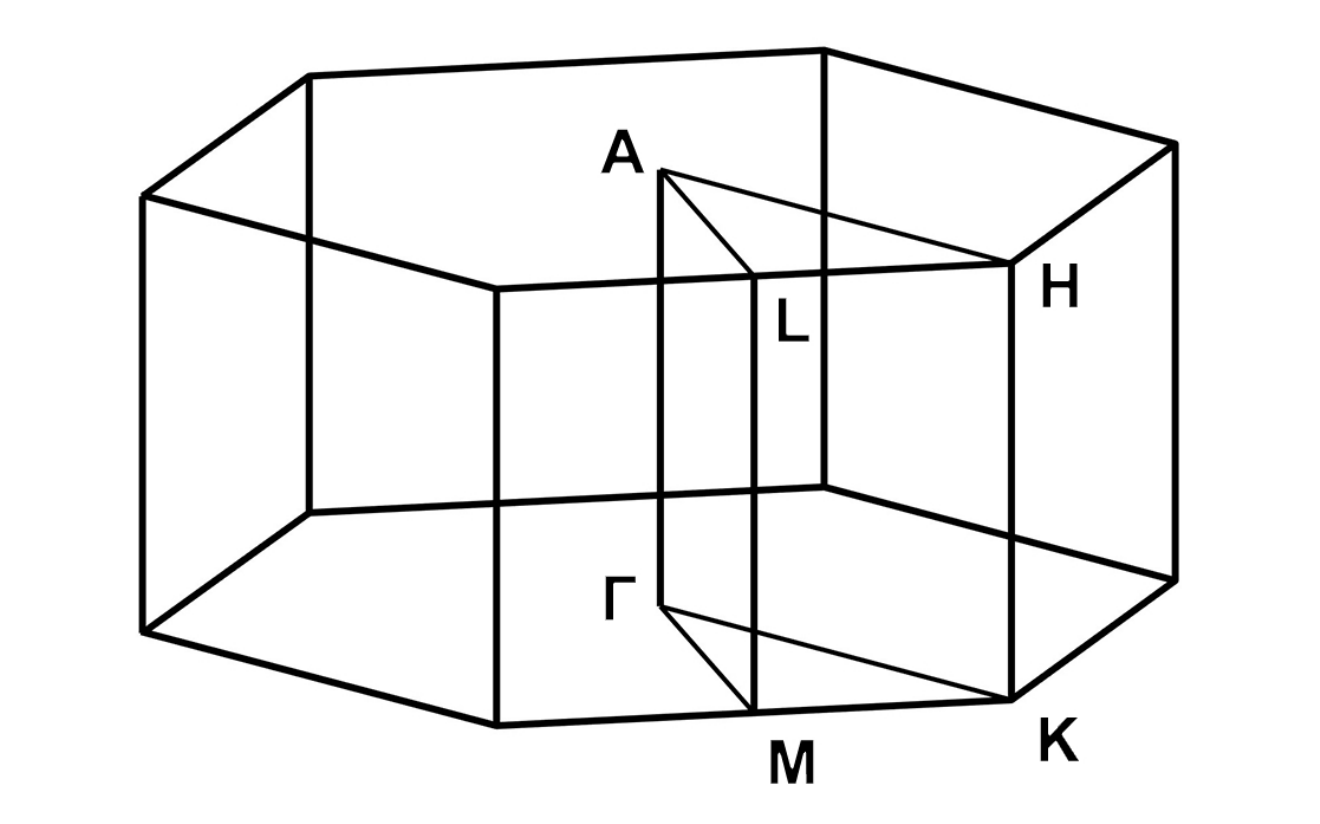}
    	\caption{3D Brillouin Zone and high-symmetry path $\Gamma$-M-K-$\Gamma$-A-L-H-A for CsV$_3$Sb$_5$.}
    	\label{fig7}
        \end{figure}
        We performed first-principle DFT calculations on CsV$_3$Sb$_5$ using the Vienna Ab initio Simulation (VASP) package \cite{kresse_efficient_1996,kresse_efficiency_1996}, with pseudo-potential based on the projector augmented wave (PAW) method \cite{blochl_projector_1994}. The exchange correlation energy was described by the generalized gradient approximation (GGA) Perdew-Burke-Ernzerhof exchange correlation functional \cite{perdew_generalized_1996}. A kinetic energy cut-off of 700 eV was applied to the plane-wave basis, which is much higher than necessary \cite{tan_charge_2021} to improve convergence. The Brillouin zone was sampled with a $\Gamma$ centered $21\times21\times13$ $\bk$ grid mesh within the Monkhorst–Pack scheme. The relaxed lattice parameters from DFT calculation is $(a,b,c)=(5.398,5.398,9.477)\text{\r{A}}$ (space group P6/mmm). Spin-orbit coupling effects were not included because it does not change the overall band structure much, consistent with a previous study on $A$V$_3$Sb$_5$ \cite{kang_twofold_2022-1,hu_rich_2022}. The 3D Brillouin zone and the high-symmetry path are demonstrated in Fig.~\ref{fig7}.
    
    \section{Slater-Koster Formalism for $d_{xy/x^2-y^2/z^2}$ Orbitals}
        \begin{figure*}
        	\centering
        	\includegraphics[width=\textwidth]{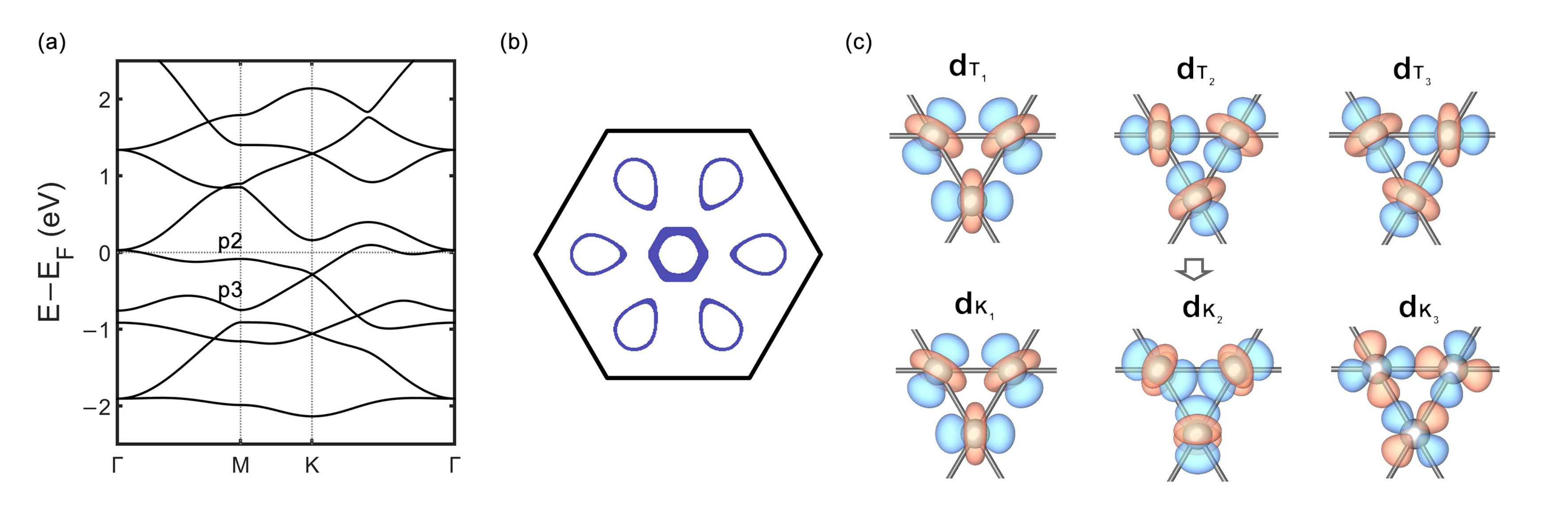}
        	\caption{SK model results for $d_{xy/x^2-y^2/z^2}$ and understanding of hybrid orbitals. (a) The band structure and (b) Fermi surfaces produced by the SK model. (c) Hybrid orbitals constructed with respect to $C_3$ rotation symmetry (first row) and combined based on mirror symmetry to get mixing angle $\theta=0.196\pi$.}
        	\label{fig9}
        \end{figure*}
        The Slater-Koster hopping functions for $d_{xy/x^2-y^2/z^2}$ orbitals with different orbital mixing angle $\theta$ between $d^r_{x^2-y^2}$ and $d^r_{z^2}$ in Eq.~\ref{hyb} are:
        \begin{widetext}
        \begin{align}
            f^{11} &= ((3t^\sigma-12t^\pi+t^\delta)/8)\times \cos(\theta)^2+((t^\sigma+3t^\delta)/2)
            \times \sin(\theta)^2-(\sqrt{3}(t^\sigma-t^\delta)/4)\times \sin(2\theta)\nonumber\\
            f^{22} &= -(9t^\sigma-4t^\pi+3t^\delta)/8\nonumber\\
            f^{33} &= ((3t^\sigma-12t^\pi+t^\delta)/8)\times \sin(\theta)^2+((t^\sigma+3t^\delta)/2)\times \cos(\theta)^2+(\sqrt{3}(t^\sigma-t^\delta)/4)\times \sin(2\theta)\nonumber\\
            f^{23} &= (\sqrt{3}(3t^\sigma+4t^\pi+t^\delta)/8)\times \sin(\theta)+(3(t^\sigma-t^\delta)/4)\times \cos(\theta)\nonumber\\
            f^{31} &= ((3t^\sigma-12t^\pi+t^\delta)/16)\times \sin(2\theta)-((t^\sigma+3t^\delta)/4)\times \sin(2\theta)+(\sqrt{3}(t^\sigma-t^\delta)/4)\times \cos(2\theta)\nonumber\\
            f^{12} &= -(\sqrt{3}(3t^\sigma+4t^\pi+t^\delta)/8)\times \cos(\theta)+(3(t^\sigma-t^\delta)/4)\times \sin(\theta)
          \end{align}
          \label{fab}
        \end{widetext}
        After fitting the obtained extended SK hopping parameters $t_\text{\nth{1}, \nth{2}}^{\sigma,\pi,\delta}$ in Table ~\ref{eparams} to these hopping functions, we can get the Slater-Koster hopping parameters $t_\text{\nth{1}}^{\sigma,\pi,\delta}$ and $t_\text{\nth{2}}^{\sigma,\pi,\delta}$, as well as the SK hopping strengths listed in Table ~\ref{skparam}. The obtained SK hopping in Table ~\ref{skparam} can be used to plot the band structure produced by the SK model in Eq.~\ref{sym_HT} without any parameter relaxation, as shown in Fig.~\ref{fig9}(a). The Fermi surfaces are shown in Fig.~\ref{fig9}(b). The energy dispersions and the Fermi surfaces of the SK model cannot describe the low-energy electronic structure of the DFT calculations in Figs.~\ref{fig2}(a) and (f). Consequently, the renormalization of the hopping strengths between the $d$ orbitals on V atoms due to other atoms is still substantial. Thus, the extended SK model is necessary to describe the electronic structure of $d_{xy/x^2-y^2/z^2}$. This further highlights the importance of understanding the multiorbital Hamiltonian to achieve a more efficient and accurate description of electronic structures, representing a meaningful step beyond parameter fine-tuning.

        \begingroup
        \renewcommand{\arraystretch}{1.5} % Default value: 1
        \begin{table}
            \centering
            \begin{tabular}{|c|c|c|c|c|c|c|c|c|c|} \hline 
                 (eV) & $f^{11}$ & $f^{22}$ & $f^{33}$ & $f^{12}$& $f^{23}$& $f^{31}$& $\mu_{1}$ & $\mu_{2}$ & $\mu_{3}$\\ \hline 
                 $f_\text{\nth{1}}$&  0.05&  -0.13&  0.23&  -0.45&  0.22& -0.34& 0.85& -0.084& 0.75\\ \hline 
                 $f_\text{\nth{2}}$&  0.14&  0.21&  -0.23&  0.02&  -0.01& 0.08& \multicolumn{3}{|c|}{}\\ \hline
            \end{tabular}
            \caption{Values of Slater-Koster functions with parameters $t_\text{\nth{1}}^\sigma=-0.189$, $t_\text{\nth{1}}^\pi=0.27$, $t_\text{\nth{1}}^\delta=0.31$ and $t_\text{\nth{2}}^\sigma=0.150$, $t_\text{\nth{2}}^\pi=-0.079$, $t_\text{\nth{2}}^\delta=0.060$.}
            \label{skparam}
        \end{table}
        \endgroup
        
    \section{Hybrid Orbital Origin}
        We can understand the origin of heavy orbital mixing between $d_{x^2-y^2}$ and $d_{z^2}$ by considering the existence of Sb atoms in the plane at the hexagonal centers. By treating these Sb atoms equally with V atoms forming the kagome lattice, we consider a triangle lattice for simplicity. First, we can find a set of $C_6$ related hybrid orbitals $d_{T_{1/2/3}}$ from a linear combination of the three $d$ orbitals on each sublattice \cite{dresselhaus2007group}. The coefficients are identical to the typical $sp^2$ hybridization. Next, group $d_{T_{1/2/3}}$ orbitals on each site following the $C_3$ symmetry with the rotation axis in the center of the triangles, as shown in the first row of Fig.~\ref{fig9}(c). The last step is to recover the symmetries of the kagome lattice by considering the mirror symmetries. By recombining the two orbitals $d_{T_2}$ and $d_{T_3}$ in mirror-symmetric and antisymmetric sets, we obtain hybrid orbitals $d_{K_{1/2/3}}$ following the symmetry of the kagome lattice as shown in the second row of Fig.~\ref{fig9}(c) with a mixing angle of $\theta=0.196\pi$, which is close to the DFT mixing angle $\theta_{DFT}=0.2236\pi$ and the extended SK TB model $\theta_{TB}=0.2041\pi$. 

        Therefore, the mixing between orbitals can be understood as a crystal field effect mainly because of the existence of Sb atoms in the center of the hexagons within the kagome lattice plane. Only with this important mixing of the $d$ orbitals can the low-energy electronic structure of the $d_{xy/x^2-y^2/z^2}$ orbitals be described.

    \section{Band Structure for $d_{xz/yz}$ orbitals}
        \begingroup
        \renewcommand{\arraystretch}{1.5} % Default value: 1
        \begin{table}
            \centering
            \begin{tabular}{|c|c|c|c|c|c|} \hline 
                 (eV) & $t^{11}$ & $t^{22}$ &$t^{12}$& $\mu_{1}$ & $\mu_{2}$ \\ \hline 
                 $t_\text{\nth{1}}$&  -0.40&  0.28&  0.45&  0.14&  -0.12\\ \hline 
                 $t_\text{\nth{2}}$&  -0.15&  -0.06&  -0.08& \multicolumn{2}{|c|}{}\\ \hline
            \end{tabular}
            \caption{Parameters of the band structure calculated by extended SK method for $d_{xz/yz}$ orbitals.}
            \label{oparams}
        \end{table}
        \endgroup
        Rotating the $d_{xz}$ and $d_{yz}$ orbitals following kagome lattice symmetry based on Eq.~\ref{rot_ob} transforms the original SK Hamiltonian to Eq.~\ref{sym_HE} \cite{hu_rich_2022,okamoto_topological_2022} in $d^r_{xz/yz}$ orbital basis:
        \begin{equation}
            \left[\begin{array}{c}
            d^{r\dagger}_{xz,i} \\
            d^{r\dagger}_{yz,i} \\
            \end{array}\right] = \left[\begin{array}{cc}
            \cos(\mathbf{\hat{r}}_{0i}) & \sin(\mathbf{\hat{r}}_{0i})\\
            -\sin(\mathbf{\hat{r}}_{0i}) & \cos(\mathbf{\hat{r}}_{0i})\\
            \end{array}\right] \left[\begin{array}{c}
            d^{\dagger}_{xz,i} \\
            d^{\dagger}_{yz,i} \\
            \end{array}\right]     
            \label{rot_ob}
        \end{equation}
        
        \begin{align}
        H^s_{o}(\mathbf{k}) &= \left[\begin{array}{cc}
        \frac{t_{\pi}-3t_{\delta}}{2}H_{11}^K+\mu_1 I & \frac{\sqrt{3}(t_{\delta}+t_{\pi})}{2}H_{12}^{AK \dagger} \nonumber\\
        \frac{\sqrt{3}(t_{\delta}+t_{\pi})}{2}H_{12}^{AK} & \frac{t_{\delta}-3t_{\pi}}{2}H_{22}^K+\mu_2 I 
        \end{array}\right] \\
        &= \left[\begin{array}{cc}
        f^{11}H_{11}^K+\mu_1 I & f^{12}H_{12}^{AK\dagger} \\
        f^{12}H_{12}^{AK} & f^{22}H_{22}^K+\mu_2 I \\
        \end{array}\right],
        \label{sym_HE}
        \end{align}
        where the interorbital hopping block has the antisymmetric structure $H^{AK}$ due to the different mirror symmetry properties (even/odd) of the $d^r_{xz/yz}$ orbitals.
        
        By relaxing the SK hopping strengths $f^{\alpha\beta}$ to independent hopping parameters $t^{\alpha\beta}$, we obtain the extended SK Hamiltonian for the $d^r_{xz/yz}$ orbitals: 
        \begin{equation}
        H_{o}(\bk) = \left[\begin{array}{cc}
        t^{11}H_{11}^K+\mu_1 I & t^{12}H_{12}^{AK\dagger} \\
        t^{12}H_{12}^{AK} & t^{22}H_{22}^K+\mu_2 I \\
        \end{array}\right]
        \label{ext_HE}
        \end{equation}
        
\begin{figure*}[ht]
        \centering
        \includegraphics[width=\textwidth]{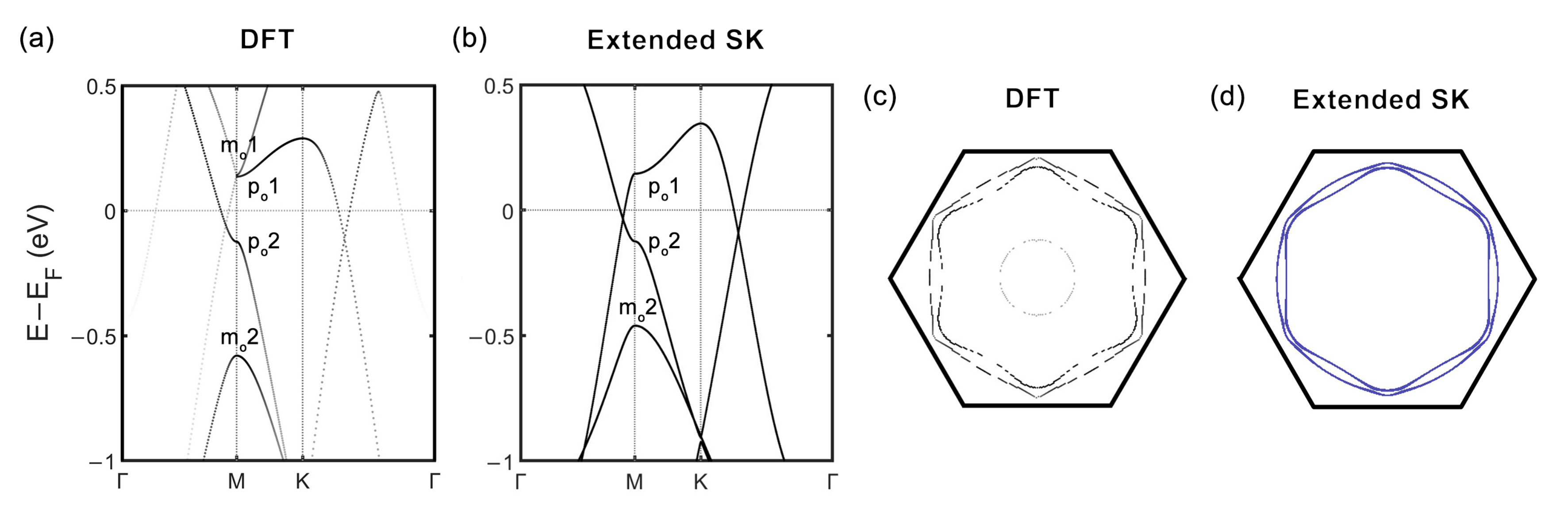}
        \caption{Comparison of DFT calculation and extended SK model for $d_{xz}$ and $d_{yz}$ orbitals. The band structure of CsV$_3$Sb$_5$ for $d_{xz}$ and $d_{yz}$ orbitals calculated by (a) DFT and (b) the extended SK model. The Fermi surfaces calculated by (c) DFT and (d) the extended SK model with parameters in Table ~\ref{oparams}.}
        \label{fig8}
        \end{figure*}

        Near the Fermi level, although the $d_{xz/yz}$ orbitals hybrid with the $p_z$ orbitals of the Sb atoms in the kagome plane and the $p_{x/y}$ orbitals of the Sb atoms outside the kagome plane \cite{jeong_crucial_2022,li_origin_2023,2023Deng}, the low-energy electronic structure can still be well captured by the extended SK TB model. To capture the crystal field effects of Sb atoms, we can also pin the p-vHS1 and p-vHS2 in Fig.~\ref{fig8}(a) at the DFT calculated energies in the extended SK model as shown in Fig.~\ref{fig8}(b).
        The energies of m-vHS1 differ for $A =$ K, Rb, Cs \cite{sim_inversion_2024}, which are not the focus of our model. 
        
        We obtain the parameters in Table ~\ref{oparams} for the band structure in Fig.~\ref{fig8} (b), which is in good agreement with the DFT calculations in Fig.~\ref{fig8}(a), where the dominant interorbital hopping strength $t^{12}_\text{\nth{1}}$ and the antisymmetric $H^{AK}$ block also play an important role. There is also great agreement between the Fermi surfaces calculated by the extended SK model in Fig.~\ref{fig8}(d) with the DFT results in Fig.~\ref{fig8}(c).
\end{document}